\documentstyle[twocolumn,eqsecnum,aps]{revtex}
\def\be{\begin{equation}}
\def\ee{\end{equation}}
\def\bea{\begin{eqnarray}}
\def\eea{\end{eqnarray}}
\begin{document}
\draft
\title{Wilson Fermions on a Transverse Lattice}
\author{M. Burkardt and H. El-Khozondar}
\address{Department of Physics\\
New Mexico State University\\
Las Cruces, NM 88003-0001\\U.S.A.}
\maketitle
\begin{abstract}
In the light-front formulation of field theory, it is possible
to write down a chirally invariant mass term. It thus appears
as if one could solve the species doubling problem on a light-front
quantized transverse lattice in a chirally invariant way. 
However, upon introducing link fields and after renormalizing, 
one finds exactly the same LF Hamiltonian as if one had started 
from the standard Wilson action in the first place. The (light-front) 
chirally invariant transverse lattice regularization is thus not
chirally invariant in the conventional sense. As an application of
the Wilson formulation for fermions on a $\perp$ lattice, we 
calculate spectrum, distribution functions and distribution amplitudes
for mesons below $2 GeV$ in a truncated Fock space.
\end{abstract}
\pacs{\\ }
\narrowtext
\section{Introduction}
There are both phenomenological as well as theoretical reasons 
to study light-front (LF) quantization: degrees of freedom that 
are close to the relevant degrees of freedom in many high energy
scattering experiments and the simplified vacuum structure.
For recent reviews on these topics see Refs. 
\cite{osu:all,brazil,world,mb:adv}.

One approach that seems particularly promising towards performing
numerical calculations of hadron bound states in $QCD_{3+1}$ is
the transverse ($\perp$) lattice \cite{bardeen}.

For our purposes, the most crucial difference between covariant field 
theories and LF field theories is the fact that LF field theories allow for
\footnote{In fact, proper renormalization requires them 
\cite{osu:all,mb:parity,mb:hala} as counter-terms.}
chirally invariant mass terms
\begin{equation}
{\cal L}_{kin} = \Delta m^2 \ \bar{\psi}\frac{\gamma^+}{2i\partial_-}\psi
\end{equation}
In the canonical quantization approach, this term arises naturally
in addition to the (chirally odd) helicity flip interaction term, which
is also proportional to the mass.

Since it is possible to write down a chirally invariant mass term, it is also
possible to write down a chirally invariant analog of the Wilson r-term
\begin{equation}
S=S_{naive} + \delta S = S_{naive}+a^2\sum_{n_x,n_y}\int dx^+dx^- \delta {\cal L}_{n_x,n_y},
\end{equation}
where $\delta {\cal L}_{n_x,n_y}$ is the lattice approximation to\\
$-ar\bar{\psi}\frac{\gamma^+}{i\partial_-}{\vec \partial}^2_\perp \psi$. 
One might thus hope that chiral fermions can be formulated on a transverse
lattice without the notorious species doubling problem \cite{privat}.
The purpose of this paper is to give an explicit example which shows that
this hope is premature. In fact, it will be shown that, after renormalization,
one recovers the standard Wilson term in the interaction.

In order to keep the complexity of the equations at a minimum,
the lattice regularization is first explained on a $\perp$ lattice
in 2+1 dimensions.
This has the advantage that one does not need to specify the direction 
in which the transverse derivative acts. After extending the formulation 
to 3+1 dimensions, we proceed to apply Wilson fermions to a simple 
$\perp$ lattice model for mesons.

\section{Fermions on a Transverse Lattice}
\label{sec:fermi}
\subsection{Naive Fermions}
We start by rewriting the Lagrangian for a free fermion (2+1 dimensions,
4-spinors for the fermions)
in terms of the LF-projections $\phi \equiv \frac{\gamma^-\gamma^+}{2}\psi$ and
$\chi \equiv \frac{\gamma^+\gamma^-}{2}\psi$
\begin{eqnarray}
{\cal L} &=& \bar{\psi}\left( i\not \!\!\;\partial -m\right)\psi
\nonumber\\
&=&\sqrt{2} \phi^\dagger i\partial_+ \phi
+\sqrt{2} \chi^\dagger i\partial_- \chi
-\phi^\dagger \left(i\alpha_x \partial_x +\beta m\right)\chi
\nonumber\\ & &-\chi^\dagger \left(i\alpha_x \partial_x +\beta m\right)\phi ,
\label{eq:cont}
\end{eqnarray}
where $\alpha_x=\gamma^0\gamma_x$ and $\beta=\gamma^0$.
Upon discretizing the transverse ($x\rightarrow n $) coordinate
and replacing transverse derivatives by finite differences in Eq. (\ref{eq:cont})
one thus finds the discretized action for (naive) free fermions on a 
transverse lattice
\begin{equation}
S_{naive}= a\sum_n \int dx^+dx^- {\cal L}_n,
\end{equation}
where (the dependence on the continuous longitudinal coordinates $x^\pm$ is
suppressed for notational convenience)
\begin{eqnarray}
{\cal L}_n&=&
\sqrt{2} \phi^\dagger _n \, i\partial_+ \, \phi_n
+ \sqrt{2} \chi^\dagger _n\, i\partial_-\, \chi_n
\nonumber\\ &-& \phi^\dagger_n\,\beta m \,\chi_n
-\chi^\dagger_n\,\beta m \,\phi _n
\nonumber\\
&-&\phi^\dagger _n i\alpha_x \frac{\chi _{n+1}-\chi _{n-1} }{2a}
-\chi^\dagger _n i\alpha_x \frac{\phi _{n+1}-\phi _{n-1} }{2a}
.
\label{eq:ldiscr}
\end{eqnarray}
Since Eq. (\ref{eq:ldiscr}) does not involve a ``time'' ($x^+$) derivative of
$\chi$, it is convenient to use the constraint equation
\begin{eqnarray}
\sqrt{2}i\partial_-\chi_n&=& \beta m \,\phi _n
+ i\alpha_x\frac{\phi _{n+1}-\phi _{n-1} }{2a}
\label{eq:constr}
\end{eqnarray}
to eliminate $\chi$ prior to quantization, yielding the effective
action for the dynamical ($\phi$) degrees of freedom
\begin{equation}
S_{naive}^{\phi}= a\sum_{n} \int dx^+dx^- {\cal L}^{\phi}_{n} ,
\end{equation}
where
\begin{eqnarray}
{\cal L}^{\phi}_{n}&=&
\sqrt{2} \phi^\dagger _n \, i\partial_+ \, \phi_n
-\phi^\dagger _n\frac{m^2}{i\sqrt{2}\partial_-}\phi_n
\label{eq:ldisc2}
\\
& &-\frac{\phi^\dagger _{n+1}-\phi^\dagger _{n-1} }{2a}
\frac{1}{i\sqrt{2}\partial_-}
\frac{\phi _{n+1}-\phi _{n-1} }{2a}
\nonumber
\end{eqnarray}
A simple plane wave ansatz shows that the above naive Lagrangian
results in a dispersion relation which exhibits species doubling
\begin{equation}
k^- = \frac{m^2 + \frac{\sin^2 (k_xa)}{a^2}}{2k^+}
\end{equation}
where $-\frac{\pi}{a} < k_x<\frac{\pi}{a}$. 
Thus, even for $a\rightarrow 0$,
modes with $k_x\approx \pi/a$ have finite energy.
Of course, this has nothing to do with the LF approach chosen here, but
is just the familiar phenomenon of species doubling for Dirac particles 
(first order derivatives!) on a lattice.

The species doubling problem on a $\perp$ lattice has been dealt with in
Ref. \cite{paul:ferm} using a generalization of the familiar Kogut-Susskind
fermions \cite{ks:ferm}. However, in this paper, we will try different
approaches, which are based on generalizations of Wilson fermions
\cite{w:ferm}.

\subsection{Ordinary Wilson Fermions} 
The most familiar ``patch'' to deal with species doubling
is to start from the naive action, plus a ``Wilson r-term''
of the form \cite{w:ferm} 
\begin{eqnarray}
\label{eq:wilsonr}
\delta {\cal L}_n &=& \frac{r}{a}\bar{\psi}_n\left[
\psi _{n+1} - 2\psi _n + \psi_{n-1}\right] \\
&=& \frac{r}{a}\left\{\phi^\dagger _n\beta
\left[ \chi_{n+1}-2\chi_n+\chi_{n-1}\right]\right.\nonumber\\
& &\left.+ \chi^\dagger _n\beta
\left[ \phi_{n+1}-2\phi_n+\phi_{n-1}\right]\right\}
\nonumber
\end{eqnarray}
where $r={\cal O}(1)$.
The well known problem with the Wilson r-term is that it has to
have the Dirac structure of a mass term in order to eliminate ``doublers''
both for particles and anti-particles and therefore chiral symmetry
is explicitly broken.

Including this Wilson r-term (\ref{eq:wilsonr}), the constraint
equation for $\chi$ reads
\bea
\sqrt{2}i\partial_- \chi _n &=& \beta m \phi _n +
i\alpha_x \frac{\phi_{n+1}-\phi_{n-1}}{2a} \nonumber\\& &-\frac{r}{a}
\beta \left[ \phi_{n+1}-2\phi_n+\phi_{n-1}\right],
\eea 
yielding for the effective Lagrangian of the dynamical
degrees of freedom ($\phi _n$)
\begin{eqnarray}
\delta {\cal L}^\phi_n &=&
\frac{2mr}{a} \phi^\dagger _n \frac{1}{i\sqrt{2}\partial_-}
\left[ \phi_{n+1}-2\phi_n+\phi_{n-1}\right]
\nonumber\\
&-&\frac{\phi^\dagger _{n+1}-2\phi^\dagger_n+\phi^\dagger _{n-1} }{a}
\frac{r^2}{i\sqrt{2}\partial_-}
\frac{\phi _{n+1}-2\phi_n+\phi _{n-1} }{a} .
\label{eq:ldisc3}
\end{eqnarray}
The dispersion relation, with the $r$-term included, reads
\begin{equation}
k^-=\frac{\left[m+\frac{2r}{a}\left(1-\cos k_xa\right)\right]^2
+\frac{\sin^2k_xa}{a^2}}{2k^+}
\label{eq:dispr}
\end{equation}
and is, by construction, free of species doubling at
$k_x\rightarrow \pm \pi/a$.

\subsection{Gauging Ordinary Wilson Fermions}
\label{sec:gauge}
An action that reduces to the QCD coupling of fermions to transverse
gluons in the
naive continuum limit is obtained by introducing link fields
$U_n$, defined on the links that connect
the site $n$ with $n+1$, using the rule
\begin{eqnarray}
\bar{\psi}_n\psi_{n+1} &\longrightarrow& \bar{\psi}_nU_n\psi_{n+1}
\nonumber\\
\bar{\psi}_{n+1}\psi_n &\longrightarrow& \bar{\psi}_{n+1}U^\dagger_n
\psi_n .
\end{eqnarray}
To keep the notation simple, we suppressed the dependence of
$U_n$ on $x^-$. The coupling to the longitudinal gluon components 
is a separate issue and does not affect the main conclusions in this section. 
We will thus omit couplings to $A^\pm$ here.
Obviously, the resulting gauge invariance includes only gauge
transformations that do not depend on $x^\pm$. However, but if one wants
to one can easily remedy this by introducing couplings to $A^\pm$ ---
which is what we will do in the model calculations in Section 
\ref{sec:femto}.

One thus finds for the action including link fields
\begin{eqnarray}
{\cal L}_n&=&
\sqrt{2} \phi^\dagger _n \, i\partial_+ \, \phi_n
+ \sqrt{2} \chi^\dagger _n\, i\partial_-\, \chi_n
\nonumber
-\phi^\dagger_n\,\beta m \,\chi_n
-\chi^\dagger_n\,\beta m \,\phi _n
\nonumber\\
& &-\phi^\dagger _n i\alpha_x \frac{U_n\chi _{n+1}-U^\dagger_{n-1}
\chi_{n-1}}{2a} \nonumber\\ 
& &-\chi^\dagger _n i\alpha_x \frac{U_n\phi _{n+1}-U^\dagger_{n-1}
\phi _{n-1} }{2a}
\label{eq:ldiscr4}
\end{eqnarray}
and
\begin{eqnarray}
\delta {\cal L}_n &=& \frac{r}{a}\bar{\psi}_n\left[
U_n\psi_{n+1} - 2\psi_{n} + U_{n-1}^\dagger\psi_{n-1}\right] \\
\label{eq:wilsonr2}
&=& \frac{r}{a}\left\{\phi^\dagger_n\beta
\left[ U_n\chi_{n+1}-2\chi_{n}+U_{n-1}^\dagger\chi_{n-1}\right]\right.
\nonumber\\
& &\left.+ \chi^\dagger_n\beta
\left[ U_n\phi_{n+1}-2\phi_n+U_{n-1}^\dagger\phi_{n-1}\right]\right\}
\nonumber
.
\end{eqnarray}
After eliminating $\chi_n$ one thus finds
\begin{equation}
{\cal L}^\phi_n= {\cal L}^{\phi,0}_n+{\cal L}^{\phi,1}_n+
{\cal L}^{\phi,2}_n,
\end{equation}
where 
\begin{equation}
{\cal L}^{\phi,0}_n=
\sqrt{2} \phi^\dagger_n \, i\partial_+ \, \phi_n
-\phi^\dagger_n\frac{M^2}{i\sqrt{2}\partial_-}\phi_n,
\end{equation}
with $M=m+2r/a$, is the term containing no link fields and
\bea
{\cal L}^{\phi,1}_n&=& \phi^\dagger_n
\frac{M}{\sqrt{2}i\partial_-}
\left\{ r \frac{\left[U_n\phi_{n+1} + U^\dagger_{n-1} \phi_{n-1} \right]}{a}
\right. \\
& &\quad \quad\quad \quad -\left.
\beta i \alpha_x \frac{\left[U_n\phi_{n+1} -U^\dagger_{n-1} \phi_{n-1}
\right]}{2a} \right\}
+ h.c. \nonumber
\eea
is the term linear in the link fields. Once one renormalizes,
the term quadratic
in the link fields [${\cal L}^{\phi,2}_n$] is determined
by the other two terms \cite{brazil,nato} \footnote{In fact, one can always
start from an un-renormalized LF-Hamiltonian without ``Compton terms''
(i.e. the terms bilinear in both fermion and boson field) and then
generate the Compton terms as counter terms.}
and thus does not need to be specified 
for the point we are trying to make in this paper.

\subsection{A Modified r-Term}
In the context of LF quantization, it is natural not to write down
the usual covariant Wilson r-term, but instead to modify only the
kinetic mass term in the Lagrangian after $\chi$ has been eliminated,
i.e. instead of Eq. (\ref{eq:wilsonr}) we introduce a modified
r-term, which is formulated directly in terms of the dynamical degrees
of freedom ($\phi_n$)
\begin{equation}
\delta \tilde{\cal L}^\phi_n= \phi^\dagger_n \frac{a r}{i\sqrt{2}\partial_-}
\frac{\phi_{n+1}-2\phi_n+\phi_{n-1}}{a^2}
\label{eq:modwilson} ,
\end{equation}
which is the discretized version of
$a r\bar{\psi}\frac{\gamma^+}{2i\partial_-}
{\partial}^2_x \psi$. This modified r-term yields a dispersion
relation 
\begin{equation}
k^-=\frac{m^2+\frac{2 r}{a}\left(1-\cos k_xa\right)
+\frac{\sin^2k_xa}{a^2}}{2k^+},
\label{eq:dispr2}
\end{equation}
which is very similar to Eq. (\ref{eq:dispr}), but 
nevertheless does not exhibit species doubling.
What makes this ``modified r-term'' especially attractive is
that it is chirally invariant (on the LF --- see Appendix).
This feature can be best seen by re-expressing Eq. (\ref{eq:modwilson})
in four component notation, yielding
\be
\delta \tilde{\cal L}^\phi_n=a r\bar{\psi}\frac{\gamma^+}{2i\partial_-}
\frac{\psi_{n+1}-2\psi_n+\psi_{n-1}}{a^2} .
\ee
At first, this seems to contradict the Nielsen-Ninomiya
theorem \cite{nogo}, which states (under very general assumptions) that
there is no chirally invariant way to solve the species doubling problem.
However,  the additional term [Eq. (\ref{eq:modwilson})]
is non-local \footnote{In the LF approach, after eliminating
constrained degrees of freedom, already the canonical
Lagrangian/Hamiltonian is non-local in the longitudinal direction.}
due to the inverse longitudinal derivative and thus the Nielsen-Ninomiya theorem does
not apply. 

In order to render Eq. (\ref{eq:modwilson}) gauge invariant,
we replace
\bea
\phi_n^\dagger \frac{1}{i\sqrt{2}\partial_-}\phi_{n+1}
&\longrightarrow&
\frac{1}{2}\phi_n^\dagger \left[
U_n\frac{1}{i\sqrt{2}\partial_-}
+\frac{1}{i\sqrt{2}\partial_-}U_n\right]
\phi_{n+1} \nonumber\\
\phi_n^\dagger \frac{1}{i\sqrt{2}\partial_-}\phi_{n-1}
&\longrightarrow&
\frac{1}{2}\phi_n^\dagger \left[
U_n\frac{1}{i\sqrt{2}\partial_-}
+\frac{1}{i\sqrt{2}\partial_-}U_n\right]
\phi_{n-1} \nonumber\\
\nonumber\\
\label{eq:minimal}
\eea
and similarly for other non-local terms. Although the gauge invariant 
extension is not unique, Eq. (\ref{eq:minimal}) represents
the minimal extension of Eq. (\ref{eq:modwilson}) that is gauge
invariant \footnote{More precisely, invariant under $x^\pm$ independent
gauge transformations. For the general case the same comments as in Section
\ref{sec:gauge} apply.}, hermitian and invariant under transverse parity
($n \rightarrow -n$). Using this rule, we find the following terms
contributing to the action based on this modified r-term 
\begin{equation}
\tilde{\cal L}^{\phi,0}_n = -\left(m^2+\frac{2 r}{a}\right)
\phi_n^\dagger \frac{1}{i\sqrt{2}\partial_-}\phi_{n},
\label{eq:l0}
\end{equation}
for the term containing no link fields
as well as
\bea
\tilde{\cal L}^{\phi,1}_n &=& 
\phi_n^\dagger 
\frac{1}{i\sqrt{2}\partial_-}
\left\{r\frac{
\left[U_n\phi_{n+1}
+U^\dagger_{n-1} \phi_{n-1}\right]}{2a}\right. \label{eq:l1}\\
& &\quad \quad \quad \quad -\left.
m\beta i \alpha_x \frac{\left[U_n\phi_{n+1} -U^\dagger_{n-1} \phi_{n-1}
\right]}{2a} \right\}
+ h.c. ,
\nonumber\eea
for the term linear in the link field. As in the previous section,
we do not exhibit the term quadratic in the link field since
it follows uniquely in the renormalization procedure.

A comparison with the results from the previous section shows
immediately that the naive r-term as well as the modified r-term
yield, up to a redefinition of parameters, 
identical expressions for
${\cal L}^{\phi,0}$, ${\cal L}^{\phi,1}$
and $\tilde{\cal L}^{\phi,0}$, $\tilde{\cal L}^{\phi,1}$. 
Since these are just bare parameters, this means that up to 
and including linear terms in the link fields, the standard
and the modified Wilson terms yield the same interaction terms.
Since the terms quadratic in the link fields are determined in
the renormalization procedure, this means that the standard
and modified  Wilson formulations on a transverse lattice
are identical.

This is the central result of this paper, since it shows that once
we introduce link fields and once we determine the ``Compton-terms''
in the LF Hamiltonian in the renormalization procedure,
also the terms quadratic in the link fields are the same for
the naive Wilson r-term and the ``modified'' Wilson r-term.
Even though our initial ansatz [Eq. (\ref{eq:modwilson})]
looked very promising, all that has been achieved is that we reproduced
the same interaction terms as the standard Wilson ansatz.
Compared to the standard Wilson ansatz, 
nothing has been gained by trying to remove species doubling using
the (LF-) chirally even kinetic mass term.

\subsection{From 2+1 to 3+1 Dimensions}
The main work in this paper so far was to show that the fermion
terms up to and including linear in the link fields are
the same for the standard and the modified Wilson formulation.
Since terms nonlocal in both transverse directions  ($x$ and $y$) 
at the same time (x and y) involve 2 link fields, these do not
occur when one focuses on terms up to and including linear.
Therefore, the 3+1 dimensional analysis turned out to closely
parallel the 2+1 dimensional analysis. Since no new physics
insight was gained --- in fact, the essential steps became
more mysterious because of the more lengthy equations ---
the details of the 3+1 dimensional analysis were omitted in this work.
The final result, i.e. equivalence between standard Wilson
and modified Wilson formulation, was the same as in 2+1 dimensions, since
coupling constant coherence between three point and four point couplings
of bosons to fermions holds in both cases.

\section{The Femtoworm Model for Mesons}
\label{sec:femto}
The coupling of quarks to the link-fields via the canonical
terms leads to helicity flip ($\gamma$-structure: ${\vec \alpha}_\perp$), 
while the r-term does not flip
the helicity of the quarks ($\gamma$-structure: $\beta $). 
In order
to investigate the ramifications of these basic features
for the interactions of quarks, let us consider mesons in the
approximation that there is at most one link field separating a
quark and an anti-quark. In this approximation, the quark and
anti-quark are always on the same site or on neighboring sites
and mesons propagate by one-boson exchange. For example, 
suppose that the quark and the anti-quark start out on the same
site and that the quark hops to a neighboring site.
Gauge invariance requires that the quark emits at the same time a 
link field quantum
on the link that is now connecting the quark and anti-quark.
In the next step the anti-quark follows behind by absorbing the 
link field quantum. Propagation over more
than one link proceeds by repeating this process.
The resulting sequence resembles the contracting and
stretching motion of an inchworm. The similarity in the
propagation mechanism and the fact that the typical length scale of the
mesons is not one inch but one Fermi, motivated the name for the model. 

The kinetic part and self-interactions of the link fields have been
introduced and discussed in Refs. \cite{bardeen,bvds} and will not be
discussed here since the focus of this paper is on the quarks here.
For details, the reader is encouraged to consult these References.

As far as the longitudinal dynamics is concerned, introducing quarks
leads to both a kinetic term
\begin{equation}
{\cal L}^{quarks}_{kin} = - \sum_{{\vec n}_\perp} 
\phi_{{\vec n}_\perp}^\dagger \frac{m^2}{i\sqrt{2}\partial_-}
\phi_{{\vec n}_\perp}
\end{equation}
as well as a contribution from the quarks to the vector current
\begin{equation}
J^+_{{\vec n}_\perp,i,j} = J^+_{{\vec n}_\perp,i,j}(U)
+ \sqrt{2} \phi_{{\vec n}_\perp,i}^\dagger \phi_{{\vec n}_\perp,j},
\end{equation}
where $i,j$ are color indices and where $J^+_{{\vec n}_\perp,i,j}(U)$
is the vector current from the link field which has been introduced
in Ref. \cite{bardeen}.
The main difference between the longitudinal gauge field coupling of
the quarks and the link-fields is that the quarks couple only to
the longitudinal gauge field at one site, while the link fields couple
to the longitudinal gauge field at the two sites that are adjacent 
to the link. In general, i.e. without truncation of the Fock space,
the physical space of gauge invariant states thus consists of closed
loops of link fields plus open strings of link fields with quarks and
anti-quarks at the end. For studying mesons in the large $N_C$ limit, 
we can focus our attention on quarks and anti-quarks connected by a string
of link fields. The main approximation in this section will be to limit
the length of this string to one lattice spacing. At first this seems
to be a rather drastic approximation. However, if one considers that
the lattice spacing in contemporary $\perp$ lattice calculations is on the
order of $0.4 fm$ \cite{bvds,mb:conf}, limiting the length of the string
to one lattice unit seems to be a reasonable approximation for a first
study of meson spectra on a $\perp$ lattice.

Since the coupling of the quark fields to the longitudinal gauge field is
local in the $\perp$ direction, the $\perp$ dynamics of the quarks is solely 
described by the hopping term. Using the result from Section \ref{sec:fermi}
one finds the following for the Dirac part of the hopping terms:
(Overall factors and the obligatory link-field operators are
suppressed for notational convenience! It is implicitly understood
that hopping is always accompanied by the emission or absorption of a 
link field quantum on the link connecting the initial and final site of
the hopping process)
\\ $\bullet$ 
naive term:

hopping in x-direction:
\begin{eqnarray}
{\cal L}^{naive}_x&=&
-\sum_{{\vec n}_\perp}
\phi_{{\vec n}_\perp}^\dagger \frac{i\beta \alpha_x}{i\partial_-}
\left(\phi_{{\vec n}_\perp+{\vec i}_x}-\phi_{{\vec n}_\perp-{\vec i}_x}\right) 
\label{eq:xhopp}
\\
&=&i\left( \frac{1}{p^+_{f}} - \frac{1}{p^+_{i}}\right)
\sum_{{\vec n}_\perp}
\left[ b^\dagger_\uparrow ({{\vec n}_\perp},p^+_f) 
b_\downarrow ({{\vec n}_\perp}+{\vec i}_x,p^+_i)\right.
\nonumber\\
& &\quad \quad \quad \quad \quad \quad \quad
\left.
- b^\dagger_\downarrow ({{\vec n}_\perp},p^+_f) 
b_\uparrow ({{\vec n}_\perp}+{\vec i}_x,p^+_i) \right.
\nonumber\\
& &\quad \quad \quad \quad \quad \quad \quad
\left. -b^\dagger_\uparrow ({{\vec n}_\perp},p^+_f) 
b_\downarrow ({{\vec n}_\perp}-{\vec i}_x,p^+_i)\right.
\nonumber\\
& &\quad \quad \quad \quad \quad \quad \quad
\left.
+ b^\dagger_\downarrow ({{\vec n}_\perp},p^+_f) 
b_\uparrow ({{\vec n}_\perp}-{\vec i}_x,p^+_i) \right.
\nonumber\\
& &\quad \quad \quad \quad \quad \quad \quad
\left. - d^\dagger_\uparrow ({{\vec n}_\perp},p^+_f) 
d_\downarrow ({{\vec n}_\perp}+{\vec i}_x,p^+_i)\right.
\nonumber\\
& &\quad \quad \quad \quad \quad \quad \quad
\left.
+ d^\dagger_\downarrow ({{\vec n}_\perp},p^+_f) 
d_\uparrow ({{\vec n}_\perp}+{\vec i}_x,p^+_i) \right.
\nonumber\\
& &\quad \quad \quad \quad \quad \quad \quad
\left. +d^\dagger_\uparrow ({{\vec n}_\perp},p^+_f) 
d_\downarrow ({{\vec n}_\perp}-{\vec i}_x,p^+_i)\right.
\nonumber\\
& &\quad \quad \quad \quad \quad \quad \quad
\left.
- d^\dagger_\downarrow ({{\vec n}_\perp},p^+_f) 
d_\uparrow ({{\vec n}_\perp}-{\vec i}_x,p^+_i)
\right] \nonumber
\end{eqnarray}
hopping in y-direction:
\begin{eqnarray}
{\cal L}^{naive}_y&=&
-\sum_{{\vec n}_\perp}
\phi_{{\vec n}_\perp}^\dagger \frac{i\beta \alpha_y}{i\partial_-}
\left(\phi_{{{\vec n}_\perp}+{\vec i}_y}-\phi_{{{\vec n}_\perp}-{\vec i}_y}\right) 
\label{eq:yhopp}
\\
&=&\left( \frac{1}{p^+_{f}} - \frac{1}{p^+_{i}}\right)
\sum_{{\vec n}_\perp}
\left[ b^\dagger_\uparrow ({{\vec n}_\perp},p^+_f) 
b_\downarrow ({{\vec n}_\perp}+{\vec i}_y,p^+_i)\right.
\nonumber\\
& &\quad \quad \quad \quad \quad \quad \quad
\left.
+ b^\dagger_\downarrow ({{\vec n}_\perp},p^+_f) 
b_\uparrow ({{\vec n}_\perp}+{\vec i}_y,p^+_i) \right.
\nonumber\\
& &\quad \quad \quad \quad \quad \quad \quad
\left. -b^\dagger_\uparrow ({{\vec n}_\perp},p^+_f) 
b_\downarrow ({{\vec n}_\perp}-{\vec i}_y,p^+_i)\right.
\nonumber\\
& &\quad \quad \quad \quad \quad \quad \quad
\left.
- b^\dagger_\downarrow ({{\vec n}_\perp},p^+_f) 
b_\uparrow ({{\vec n}_\perp}-{\vec i}_y,p^+_i) \right.
\nonumber\\
& &\quad \quad \quad \quad \quad \quad \quad
\left. - d^\dagger_\uparrow ({{\vec n}_\perp},p^+_f) 
d_\downarrow ({{\vec n}_\perp}+{\vec i}_y,p^+_i)\right.
\nonumber\\
& &\quad \quad \quad \quad \quad \quad \quad
\left.
- d^\dagger_\downarrow ({{\vec n}_\perp},p^+_f) 
d_\uparrow ({{\vec n}_\perp}+{\vec i}_y,p^+_i) \right.
\nonumber\\
& &\quad \quad \quad \quad \quad \quad \quad
\left. +d^\dagger_\uparrow ({{\vec n}_\perp},p^+_f) 
d_\downarrow ({{\vec n}_\perp}-{\vec i}_y,p^+_i)\right.
\nonumber\\
& &\quad \quad \quad \quad \quad \quad \quad
\left.
+ d^\dagger_\downarrow ({{\vec n}_\perp},p^+_f) 
d_\uparrow ({{\vec n}_\perp}-{\vec i}_y,p^+_i)
\right] \nonumber
\end{eqnarray}
$\bullet$ r-term:
\begin{eqnarray}
& &\label{eq:rhopp}\\
{\cal L}^{r}&=&\sum_{{\vec n}_\perp}
\phi_{{\vec n}_\perp}^\dagger \frac{1}{i\partial_-}
\left(
\phi_{{{\vec n}_\perp}+{\vec i}_x}+
\phi_{{{\vec n}_\perp}-{\vec i}_x}+
\phi_{{{\vec n}_\perp}+{\vec i}_y}+
\phi_{{{\vec n}_\perp}-{\vec i}_y}
\right) \nonumber
\\
&=& \left( \frac{1}{p^+_{f}} + \frac{1}{p^+_{i}}\right)
\sum_{{\vec n}_\perp}
\left[ b^\dagger_\uparrow ({{\vec n}_\perp},p^+_f) 
b_\uparrow ({{\vec n}_\perp}+{\vec i}_x,p^+_i)\right.
\nonumber\\
& &\quad \quad \quad \quad \quad \quad \quad
\left.
+ b^\dagger_\downarrow ({{\vec n}_\perp},p^+_f) 
b_\downarrow ({{\vec n}_\perp}+{\vec i}_x,p^+_i) \right.
\nonumber\\
& &\quad \quad \quad \quad \quad \quad \quad
\left. +d^\dagger_\uparrow ({{\vec n}_\perp},p^+_f) 
d_\uparrow ({{\vec n}_\perp}-{\vec i}_x,p^+_i)\right.
\nonumber\\
& &\quad \quad \quad \quad \quad \quad \quad
\left.
+ d^\dagger_\downarrow ({{\vec n}_\perp},p^+_f) 
d_\downarrow ({{\vec n}_\perp}-{\vec i}_x,p^+_i) \right.
\nonumber\\
& &\quad \quad \quad \quad \quad \quad \quad
\left.
+b^\dagger_\uparrow ({{\vec n}_\perp},p^+_f) 
b_\uparrow ({{\vec n}_\perp}+{\vec i}_y,p^+_i)\right.
\nonumber\\
& &\quad \quad \quad \quad \quad \quad \quad
\left.
+ b^\dagger_\downarrow ({{\vec n}_\perp},p^+_f) 
b_\downarrow ({{\vec n}_\perp}+{\vec i}_y,p^+_i) \right.
\nonumber\\
& &\quad \quad \quad \quad \quad \quad \quad
\left. +d^\dagger_\uparrow ({{\vec n}_\perp},p^+_f) 
d_\uparrow ({{\vec n}_\perp}-{\vec i}_y,p^+_i)\right.
\nonumber\\
& &\quad \quad \quad \quad \quad \quad \quad
\left.
+ d^\dagger_\downarrow ({{\vec n}_\perp},p^+_f) 
d_\downarrow ({{\vec n}_\perp}-{\vec i}_y,p^+_i)
\right] \nonumber
\end{eqnarray}
Note that in the naive term, hopping in the the positive $x$ or $y$
direction ($n \rightarrow n+1$) yields the opposite sign from hopping
in the negative direction ($n \rightarrow n-1$). In the r-term, 
the sign is the same for hopping in all directions. This observation will 
become important when we discuss interference between various contributions.

\subsection{Solutions with ${\vec P}_\perp^{total}=0$.}
\label{sec:p0}
Because of manifest (discrete) translational invariance in the
$\perp$ direction, the total transverse momentum 
${\vec P}_\perp^{total}$ is conserved. As we will see below, the
case ${\vec P}_\perp^{total}=0$ yields particularly simple
bound state equations and in the following we will focus our attention
on this case.

The first simplification is a non-interference between naive hopping and 
r-term hopping: In general, one cannot distinguish between a link quantum 
emitted by the naive term and one that was emitted from the r-term. Thus
a link quantum emitted from the naive term when the quark hops by one
lattice unit may be absorbed by the r-term when the anti-quark follows 
behind and vice versa. Since the naive term flips the spin, but the
r-term does not, this interference results in a non-conservation of the
meson spin (which is of course a lattice artifact). However, this 
interference term has opposite signs for hopping into opposite 
directions. Therefore, for ${\vec P}^{total}_\perp =0$, there is an 
exact cancellation
between interference terms when the meson hops in opposite directions
and there is not interference between link quanta emitted from the
naive term and those emitted from the r-term.

Note that, since the naive term flips quark helicity and the r-term
does not, the above result has an important consequence for the 
spin structure of the $q\bar{q}$ (2 particle) Fock component:
Because of the non-interference mentioned above, the only conceivable
mixing of 2-particle Fock components could be between states where
$q$ and $\bar{q}$ have both spins up or both spins down, i.e.
``double helicity flip'' . However, since the spin matrix element
for hopping in the x-direction with double helicity flip is 
$(-i)(+i)=1$ (\ref{eq:xhopp}), while
the same matrix element for $y$-hopping is $(+1)(-1)=-1$ 
(\ref{eq:yhopp}), both terms cancel for a plane wave with
${\vec P}_\perp=0$ and there is no double helicity flip
hopping. As a result, the total spin in the 2-particle
Fock component is conserved.

From a more general point of view, the above result about $S_z$ conservation
in the 2-particle Fock component emerges since the
transverse lattice is invariant under rotations by $\pi /2$ around
the z-axis, plane wave solutions with ${\vec P}_\perp^{total}=0$ must
be eigenstates under rotations around the z-axis by $\pi /2$ (with
eigenvalues $\pm 1$ and $\pm i$, i.e. in a sense $J_z$ is conserved
modulo 4).
Since there is no orbital or gluon field angular momentum when the 
$q\bar{q}$ pair is on the same site this means that $S_z$, which can
assume only the values $0$ and $\pm 1$, must be conserved.

Another important observation concerns the sign of the spin splittings
between mesons. For $P_x=0$ one finds that x-hopping
is attractive in the pseudoscalar channel 
(spin wave function: $\uparrow \downarrow - \downarrow \uparrow$), since for example
$b(\uparrow \rightarrow \downarrow) = -i$ and
$d(\downarrow \rightarrow \uparrow) = -i$, i.e. together with 
another $(-)$ from the spin wave function and another sign from the
energy denominator the total sign is $(-)$, i.e.
attractive. The same result holds for
y-hopping ($k_P=0$), since for example
$b(\uparrow \rightarrow \downarrow) = 1$ and
$d(\downarrow \rightarrow \uparrow) = -1$.
Likewise one can show that naive hopping yields a repulsive interaction
for the vector channel (spin wave function: 
$\uparrow \downarrow + \downarrow \uparrow$), i.e. despite all approximations,
the Weingarten inequality \cite{wein} is satisfied for our model.

\subsection{Pure $\bar{q}Uq$ States}
\label{sec:quq}
In the previous section we have seen that (discrete) rotational invariance
yields a drastic simplification in the spin structure for
${\vec P}_\perp^{total}=0$ solutions to our model.
Again for ${\vec P}_\perp^{total}=0$, we will demonstrate in this section 
that there are some particularly simple solutions which have a vanishing
2-particle Fock component.

Mesons which mix with $q\bar{q}$ states at the same site must
necessarily have total helicity \footnote{We define total helicity 
through the eigenvalues of the discrete rotation operator for rotations
by $\pi/2$ around the z-axis. Since the resulting definition is unique 
only up to four times an integer, we break the ambiguity by always
selecting the solution with $|h|\leq 2$.}
$h=0,\pm 1$. However, even in the one link
approximation, it is very easy to construct
states which cannot mix with a $q\bar{q}$ state at the same site.
The generic form of such states is
\begin{eqnarray}
|h=2\rangle &=&\sum_{{\vec n}_\perp}\left[
b^\dagger_\uparrow ({\vec n}_\perp + {\vec i}_x)
d^\dagger_\uparrow ({\vec n}_\perp )
+i b^\dagger_\uparrow ({\vec n}_\perp + {\vec i}_y)
d^\dagger_\uparrow ({\vec n}_\perp )
\right. \nonumber\\
& &-\left. b^\dagger_\uparrow ({\vec n}_\perp - {\vec i}_x)
d^\dagger_\uparrow ({\vec n}_\perp )
-  i b^\dagger_\uparrow ({\vec n}_\perp - {\vec i}_y)
d^\dagger_\uparrow ({\vec n}_\perp )
\right]|0\rangle 
\nonumber\\
\label{eq:h2}
\end{eqnarray}
and similarly for $h=-2$
with the longitudinal wave function ($k^+$-dependence) factorized
and with the link-field implicitly included.
The angular dependence in Eq. (\ref{eq:h2}) resembles that of a state
with $L_z=+1$, but strictly speaking on a lattice there is of course no
rotational symmetry and hence no angular momentum.
Nevertheless, a state with orbital structure as in Eq. 
(\ref{eq:h2}) cannot mix with states where a $q\bar{q}$ pair is
at the same transverse site: first we note that the matrix element
for hopping due to the $r$-term is independent of the direction 
and thus there is destructive interference
when adding up the phases from the state for the quark
to hop from different directions to the site of the anti-quark
($1+i-1-i=0$). The phase of the matrix elements of the naive term for a quark
with positive helicity to hop to the site of the anti-quark are
$-i, 1, i, -1$ for hopping from the $+x, +y, -x,-y$ direction respectively.
Together with the phase of the wave function, one thus obtains 
destructive interference for the amplitude due to the r-term as well
\bea
\langle \bar{q}q|H_{q-hop}|h=2 \rangle &\propto&
(-i)(+1)+(+1)(+i)\nonumber\\ & &+
(+i)(-1)+(-1)(-i)=0 .
\eea
Similarly, one can verify that the total amplitude for anti-quark
hopping is zero for the $h=2$ states (\ref{eq:h2}).
Note that it was crucial for the cancellation argument that
the helicity of both $q$ and $\bar{q}$ is positive. 

There are more states that do not mix with
$\bar{q}q$ states for ${\vec P}_\perp^{total}=0$.
For example, any state with $L_z=2$, regardless of spin, i.e. any state of the
form
\begin{eqnarray}
|L_z=2\rangle =
& &\sum_{{\vec n}_\perp}\left[
b^\dagger ({\vec n}_\perp + {\vec i}_x)
d^\dagger ({\vec n}_\perp )
- b^\dagger ({\vec n}_\perp + {\vec i}_y)
d^\dagger ({\vec n}_\perp )\right.
\nonumber\\ & & 
+b^\dagger ({\vec n}_\perp - {\vec i}_x)
d^\dagger ({\vec n}_\perp ) 
- \left.  b^\dagger ({\vec n}_\perp - {\vec i}_y)
d^\dagger ({\vec n}_\perp )
\right]|0\rangle \nonumber\\ \label{eq:l2}
\end{eqnarray}
has a vanishing matrix element with states where the $\bar{q}q$ state is on
the same site.
To see this, let us first consider r-term hopping. Since there are no phases
for hopping in different directions from this term in the Hamiltonian, 
the only phases arise from the state and the add up to zero
$1-1+1-1=0$. This argument applies regardless of spin.
The matrix elements of the naive hopping term between Eq. (\ref{eq:l2}) and
a pure $\bar{q}q$ state vanish, since the naive hopping term in the 
Hamiltonian has always opposite signs for hopping in opposite directions,
the wavefunction yields the same sign for separations in opposite
directions. The total matrix element of the naive hopping term is thus
zero due to cancellation from opposite directions.
The pure $\bar{q}Gq$ states are summarized in Table I. 

\begin{table}
\label{table1}
\caption{Quantum numbers of states that cannot mix with $\bar{q}q$ states.
The C and P quantum numbers contain all signs from intrinsic parities, spin
and transverse symmetry but do not include signs from the longitudinal
orbital wave function.}
\begin{tabular}{|c|c|c|c|c|c|} \hline
helicity & $L_z$$\quad$ & $S$ $\quad$& $S_z$$\quad$ & C$\quad$ & P $\quad$
\\ \hline
2 & 1$\quad$ & 1$\quad$ & 1$\quad$ & +$\quad$ & +$\quad$\\
2 & 2$\quad$ & 1$\quad$ & 0$\quad$ & -$\quad$ & -$\quad$\\
2 & 2$\quad$ & 0$\quad$ & 0$\quad$ & +$\quad$ & -$\quad$\\
1 & 2$\quad$ & 1$\quad$ & -1$\quad$ & -$\quad$ & -$\quad$\\
\hline
\end{tabular}
\end{table}

The total helicity of these states is obtained from the sum of the 
discrete orbital angular momentum and the spin (which is conserved modulo 4). 
The parity and C-parity quantum numbers given in Table I
include all intrinsic
parities as well as signs from the spin and transverse wave functions.
For a state with no nodes in the longitudinal wave function, the P and
C parities given in the table are thus immediately equal to the P and C
parity of the corresponding state. However, for more complicated 
longitudinal wave functions, the P and C parities in the table need to 
be multiplied by the signs from the longitudinal wave function.

The longitudinal dynamics of all the states in Table I
are described by the same relatively 
simple wave equation, which reflects the fact that (note 
$N_C \rightarrow \infty$) the quark and anti-quark interact only with the
link field via the 1+1 dimensional gauge interaction
\bea
\!& &\!\!\!\!\!\!M^2 \psi (x,y) = \left[
\frac{m_q^2-G^2}{x} + \frac{m_{\bar{q}}^2-G^2}{y} + 
\frac{m_U^2}{1-x-y}\right] \psi (x,y)
\nonumber
\\ 
&-& G^2 \!\!\int_0^{1-y}\!\!\!\! \!\!\!\!\frac{dz}{(x-z)^2}\!\! \!
\frac{(1-x-y) + (1-z-y)}{2\sqrt{1-x-y}\sqrt{1-z-y}}
\psi (z,y)
\nonumber\\
&-& G^2 \!\!\int_0^{1-x}\!\!\!\! \!\!\!\!\frac{dz}{(y-z)^2}\!\!\! 
\frac{(1-x-y) + (1-z-y)}{2\sqrt{1-x-y}\sqrt{1-z-y}}
\psi (z,y)
\label{eq:quq}
\eea
where $x,y$ are the longitudinal momentum fractions of the quark and
anti-quark respectively and where $G^2 = \frac{g^2N}{2\pi}$.
The integrals in Eq. (\ref{eq:quq}) are principal value integrals.

Note that the link fields should {\it not} be directly interpreted as gluons
and therefore the mesons in Table I, i.e. the solutions to Eq. (\ref{eq:quq}),
are {\it not} necessarily
hybrid or exotic mesons even though they cannot mix with pure 
$\bar{q}q$ states.
The physical reason why these states cannot mix with $\bar{q}q$ states
on the transverse lattice is that conservation laws (e.g. angular momentum)
do not allow that $q$ and $\bar{q}$ sit on the same transverse coordinate,
and gauge invariance on the $\perp$ lattice requires the presence of link 
fields. However, hybrid mesons correspond to those solutions to
Eq. (\ref{eq:quq}) where the link field is in an excited state.

\subsection{Numerical Calculations}
In the numerical calculations, we solved the coupled Fock space equations,
including $\bar{q}q$ and $\bar{q}Uq$ states projected onto vanishing total
transverse momentum ${\vec P}_\perp^{total} =0$.

Furthermore, after making this plane wave ansatz, one can reduce the
equations of motion to equations that resemble the coupled Fock space
equations in collinear QCD (with truncation to 3 particles) except
that the link fields not only have vector couplings to the quarks but also
scalar couplings from the r-term. Since collinear QCD models have been
studied extensively in the literature, we omit these details here
and refer the interested reader to Refs. \cite{anton,mb:mesons}.

In the Hamiltonian, we dropped the seagull interactions involving
instantaneous fermion exchange! As has been explained in Ref. 
\cite{mb:seoul,mb:chiral}, 
gluon loops connected to gluon emission and absorption
vertices are important for dynamical vertex mass generation. If one
works with a truncated Fock space, such vertex corrections are included in 
an asymmetric way: for example, if one limits the Fock space to
$\bar{q}q$ and $\bar{q}Uq$ then vertex corrections to boson emission
can only be generated by the Hamiltonian in the ``in'' state but
not in the ``out'' state because the ``out'' state contains already
the emitted boson (and conversely for absorption). As a result, in
and out states are treated asymmetrically.
In order to understand the consequences of such an asymmetric treatment,
consider the momentum dependence of the bare emission vertex
(helicity flip)
\be
T_{emission} \propto \left( \frac{m_q}{p_i^+} -  \frac{m_q}{p_f^+}\right)
\frac{1}{\sqrt{k_U^+}} .
\ee
Loop corrections in the initial state
connected to the vertex by instantaneous interactions renormalize only
the mass term which gets divided by $p_i^+$ but not the mass term
divided by $p_f^+$ (and vice versa for loop corrections in the final state)
\cite{mb:chiral}.
However, this is exactly what happens to an emission vertex when the 
Hamiltonian is 
diagonalized in a Fock space truncated basis. There are several possible
remedies one might think of, such as perturbatively including
loop corrections which are suppressed by the Fock space truncation. 
\footnote{Note that DLCQ does {\it not} completely solve this problem, even
if all Fock components allowed by the discretization are included, because
of the loss of boost invariance in subgraphs.}
However, in this paper we chose the most simple solution and dropped
instantaneous fermion interactions altogether, which removes
loop corrections connected to vertices by instantaneous interactions 
completely. Since we are aware that those interactions are crucial for 
dynamical vertex mass generation \cite{mb:seoul,mb:chiral}, 
we compensate for this omission by using
constituent quark masses (300MeV) for the vertex mass and not current masses.
Note that since there is also an unspecified 
vertex coupling constant multiplying
the link filed emission terms, the precise numerical value of the vertex mass
is not crucial. However, it is important to note that we used a value that
does {\it not} vanish in the chiral limit. Furthermore, it is important 
that, for simplicity, we do not allow this coupling to run \cite{mb:chiral}.

Furthermore, in the $\bar{q}Uq$ Fock component, we use a constituent mass
for the kinetic mass of the quarks as well.
In the $\bar{q}q$ component of the Fock space, where one has to add
an infinite counter-term for the one loop self energy, we use the finite part
of the kinetic mass as a fit parameter to fit the $\pi$ and $\rho$ masses.

In general, the self interactions of the link field are described by a
complicated effective potential \cite{bvds}. However, in the one link
approximation only the quadratic term (the mass term) 
contributes, which we take to be about
$300MeV$ as well, which lies in the range of values suggested in Ref. 
\cite{bvds}. In the spirit of an effective link field interaction (obtained
in principle by ``smearing'' and integrating out short range fluctuations)
it would be consistent to also introduce effective interactions for the
quark fields. For example, an effective interaction induced by short range
fluctuations could be a spin dependent ``contact interaction'' 
similar to Gross-Neveu interactions. It is also conceivable that one could
introduce a spin dependent contact term such that it partially cancels the
violations of rotational invariance.
However, in this first study case, we wanted to explore how well one can
reproduce the light meson spectrum with canonical terms only and we considered
canonical quark terms only.

The most important scale in  $\perp$ lattice calculations is set by the
dimensionful coupling of the quark and link fields to the
instantaneous Coulomb interaction acting on each site.
If one suppresses higher Fock components then this coupling directly 
translates into the longitudinal string tension ($\sigma = \frac{1}{2}G^2$, 
which we fix at $\sigma =1 GeV/fm$. 
We have also tried smaller values of the string tension, but then we 
were forced to use a very large coupling for the hopping term
in order to generate the experimentally observed $\pi-\rho$ splitting
and the spectrum became numerically unstable. This is because,
for a given quark mass, smaller string tensions lead to wave functions which 
vanish faster at the kinematic end-points and hence lead to smaller 
contributions from the one boson exchange interaction which is also singular 
at these end-points.

In our numerical 
calculations we applied a sharp cutoff to the matrix element for hopping 
(i.e. link field emission and absorption) processes if the momentum of 
the final state quark is smaller than a fraction $\varepsilon$ of the 
momentum of the initial state quark. We varied this cutoff between 
$\varepsilon = 0.2$
and $\varepsilon = 0.1$ and verified that the spectra are stable
with respect to variations of this momentum fraction cutoff.

As long as the r-term has a coupling which is about $1/4$ of the coupling
for the naive hopping  term or smaller, the low energy 
spectrum for ${\vec P}_\perp =0$ is rather insensitive to its precise value. 
In fact, for such values, the main effect of the r-term is to raise the
energy of mesons consisting of ``doubler quarks'' from the edge of the 
Brillouin zone. When the r-term 
was taken to be larger (about $1/2$ relative to the naive hopping
term or more) there was also a strong effect on the low energy spectrum 
for ${\vec P}_\perp=0$ and we were no longer able to simultaneously fit the
$\pi$ and $\rho$ meson masses. We thus performed the numerical calculations 
where the value of the r-term coefficient is $1/4$ of the naive term 
coefficient.

Given all constraints listed above, we fitted the only remaining free 
parameters --- the finite part of the kinetic mass counter-term and
the vertex mass times the coupling of quarks to ${\vec A}_\perp$  ---
to the $\pi$ mass as well as the $\rho$ mass. Since rotational invariance and
parity are not manifest symmetries, it is necessary to be more specific here
about what we mean by the $\pi$ and the $\rho$ mesons.
The $\pi$-meson is taken to be the lightest meson with $C=+$ and helicity $h=0$.
Since the $\rho$ meson has spin $1$, it can have both helicity $h=0$ as well as
$h=\pm 1$. Since our approach as well as model approximations break rotational
invariance, we do not find these states to be degenerate: the lightest meson 
with $h=1$ and $C=-$ and the lightest meson with $h=0$ and $C=-$ are not exactly 
degenerate. In the fitting procedure, we chose to fit the center of mass of these
two solutions to the experimentally observed $\rho$ meson mass 
$m_\rho=770 MeV$.

\begin{figure}
\unitlength1.cm
\begin{picture}(15,13)(2,3.5)
\includegraphics{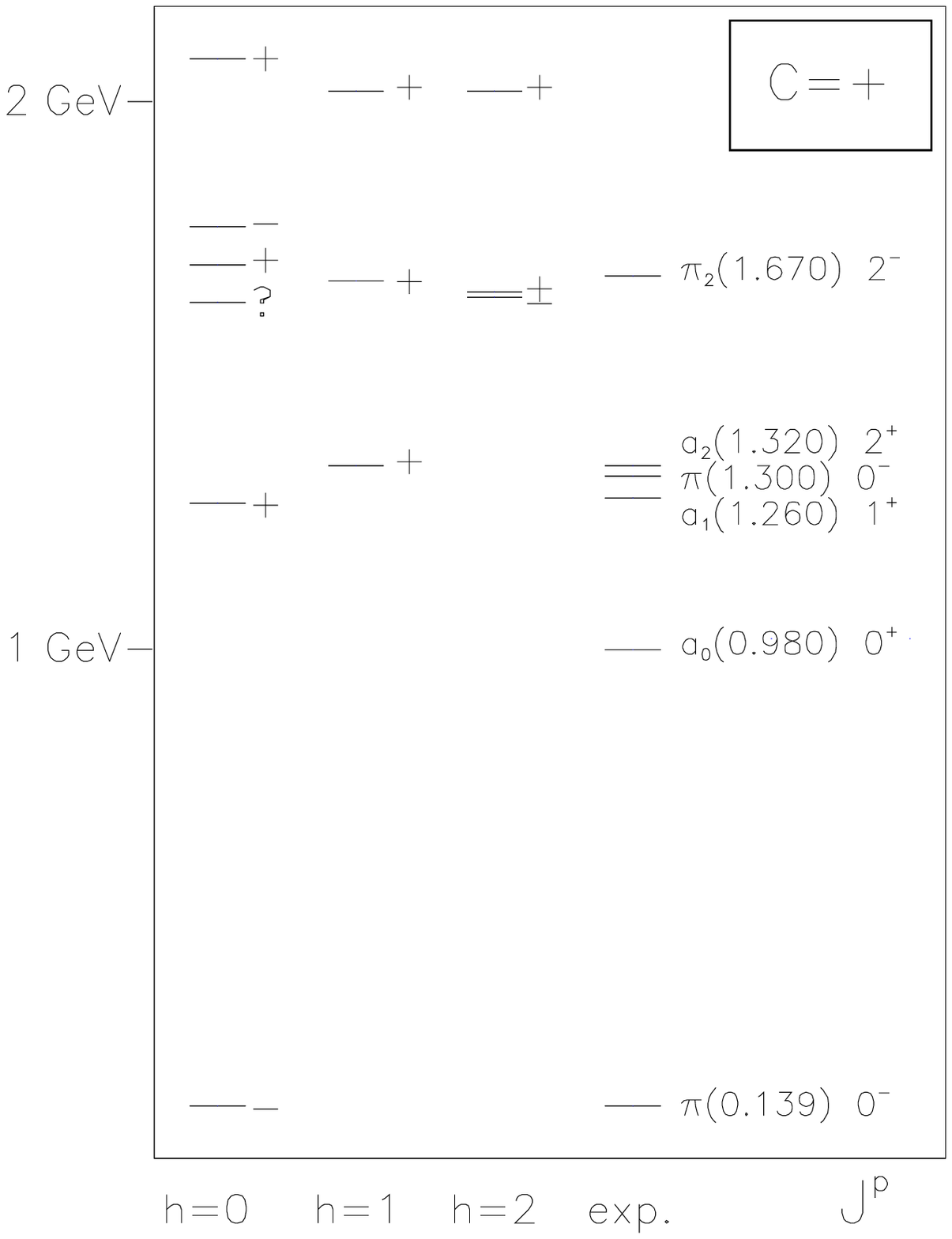}
\end{picture}
\caption{Comparison of the calculated meson spectrum for helicities
$h=0,1,2$ with the experimental meson spectrum for states with
positive C-parity. Those states where an obvious assignment for the
parity quantum number seemed possible are labelled with a `+' or `-',
while states with ambiguous parity quantum number are labels with a
`?'. Degenerate states are displaced slightly in the vertical
direction in order to make the degeneracy visible.}
\label{fig:cplus}
\end{figure}

\begin{figure}
\unitlength1.cm
\begin{picture}(15,13)(2,3.5)
\includegraphics{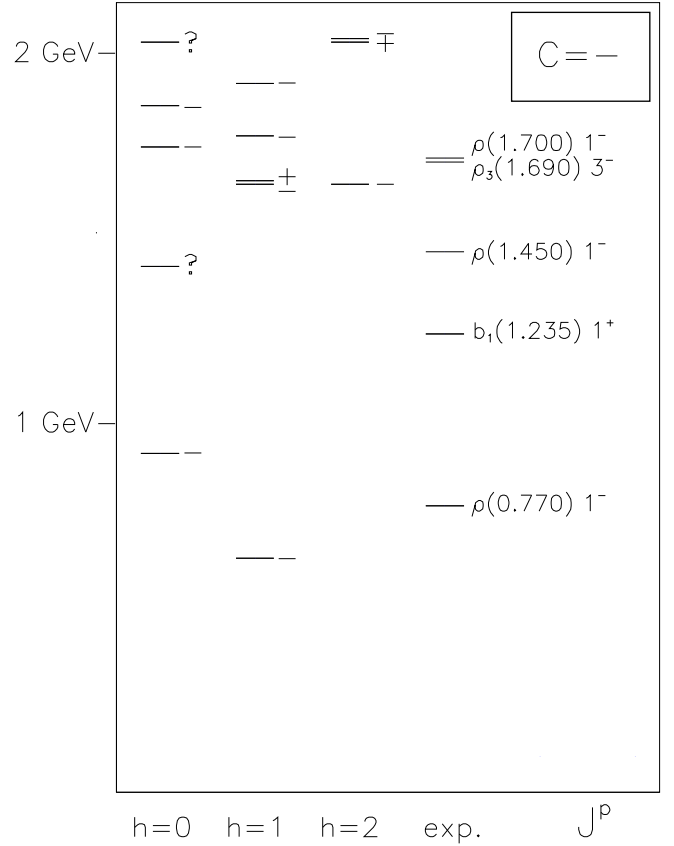}
\end{picture}
\caption{Same as Fig. \ref{fig:cplus}, but for negative C-parity.}
\label{fig:cminus}
\end{figure}

The resulting meson spectrum is shown in Fig. \ref{fig:cplus} for mesons with
positive C-parity and in Fig. \ref{fig:cminus} for mesons with negative 
C-parity.
Unlike C-parity, the usual parity is not a manifest symmetry
in the LF framework. For states with a large 2-particle component,
we used the symmetry of the 2-particle component to assign a parity quantum
number: for example, let us consider the vector current, which can
obviously only couple to negative parity states. The vacuum to meson
matrix elements of the good component $\bar{\psi} \gamma^+ \psi$ of the 
vector current operator is proportional to $\int_0^1 dx \phi_n(x)$, where
$\phi_n(x)$ is the 2-particle component of the wave function.
Obviously, this matrix element can be non-vanishing only if the 2-particle
component of the wave function is even under $x \leftrightarrow (1-x)$.
Similar observations can be made for positive C-parity operators.
We therefore assign a negative parity to states with an even wave function
in the 2-particle component of the Fock space, and a positive parity to
states with an odd wave function in the 2-particle component.
Note that this assignment is also consistent with simple quark model
considerations (i.e. free particle parity).

For states with very small or vanishing 2-particle component, this procedure
becomes dubious or useless. In those cases we determine the parity of
a state from the symmetry of the 3-particle wave function:
First there is transverse parity (which is a manifest symmetry both in 
LF quantization and on the $\perp$ lattice).
Secondly, for the ``longitudinal parity'' we combine the intrinsic parity
of a $q\bar{q}$ pair $(-)$ \footnote{Note that there is no intrinsic 
parity for the link field.}
with the orbital parity. The orbital parity
is determined using the free (no interactions) parity operator.
Note that instead of applying the free parity operator, one can simply count
nodal lines: in addition to transverse and intrinsic parity, there is a $(-)$ 
if the 3-particle wave function is odd under $q\bar{q}$ exchange and there is
another $(-)$ if there is an odd number of nodes in the wavefunction 
describing the motion of the link field relative to the ``center of mass''
of the $q\bar{q}$ pair. For example, a state with positive transverse
parity and where the longitudinal $\bar{q}Uq$ wave 
function has no node at all would be assigned $P=-$.
Of course, counting of nodes is sometimes
ambiguous, but in those cases using the free parity operator is also
not very conclusive.

However, while the above parity assignments procedure seemed reasonable to us,
it is not completely unambiguous and we thus plotted states with positive and
negative parities in the same figure.

There are several observations one can make from Figs. \ref{fig:cplus}
and \ref{fig:cminus}. First of all, because of the violations of rotational
invariance, states that one would expect to be degenerate (such as the 
h=0 and h=1 states of the $\rho$ meson) are not degenerate. However,
considering the crudeness of the model, the violations of rotational
invariance are actually moderate. The second observation one can make is
that even though our model properly describes the gross features of the
spectrum, there are clearly states ``missing'' in our model and the
energies of excited states typically lie too high. The missing states
can be easily understood by noting that the truncation of the Fock space
to 3 particles suppresses degrees of freedom associated with higher Fock
components. The result that excited states lie systematically too high
also has to do with the fact that a lot of physics has been left out by our
truncation of the Fock space. For example, restricting the maximum
separation between quark and anti-quark to one link, has a stronger
effect on excited states since those tend to be larger (if one allows them
to spread out).

\begin{figure}
\unitlength1.cm
\begin{picture}(15,11)(1,3)
\includegraphics{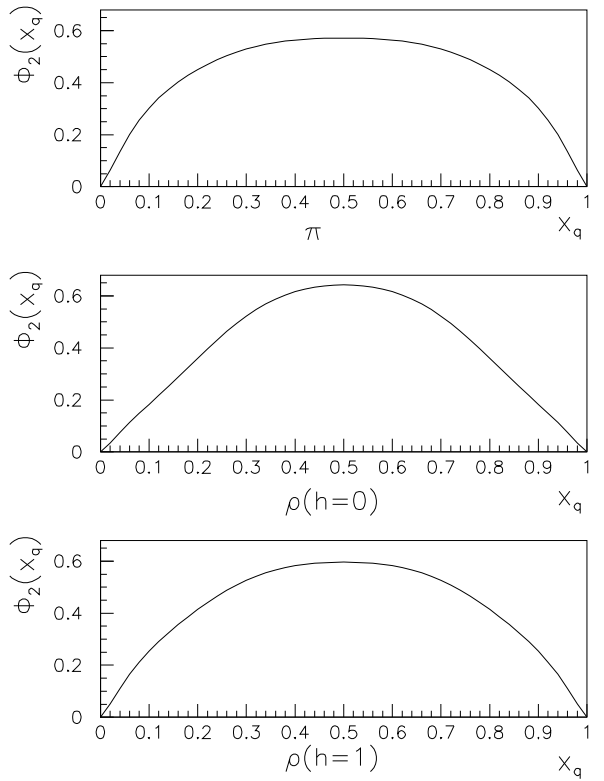}
\end{picture}
\caption{LF wave functions $\phi_2(x_q)$ in the 2 particle component
of the Fock space versus the momentum fraction carried by the quark $x_q$
for the $\pi$, and the $h=0,1$ states of the 
$\rho$ meson.}
\label{fig:wavef}
\end{figure}

Encouraged by the numerical spectrum, we
proceeded to calculate other hadronic observables --- particularly those
where the use of LF field theory is advantageous: LF wave functions
and parton distribution functions. LF wave functions or distribution
functions \footnote{These are defined to be the longitudinal momentum space
LF wave function (where momenta are measured in units of the total momentum)
in the $q\bar{q}$ Fock component with zero transverse 
separation, i.e in our calculation the valence wave function.}
for the $\pi$ and the two helicity states for the $\rho$ are shown in Fig.
\ref{fig:wavef}
What one observes in Fig. \ref{fig:wavef} is that 
all of them are rather flat with a maximum in the middle, which reflects
the strong binding of these states. Thus despite the low scale
(recall our large transverse lattice spacing $a\approx .4 fm$)
the pion wave function $\phi_\pi(x)$ has a shape which already resembles
the asymptotic form $\phi_\pi (x) \propto x(1-x)$, but it is much more flat.
A closer look at the wave functions shows that the $\rho$ meson wave functions
are slightly more peaked in the middle than $\phi_\pi$ which reflects the
weaker binding of the $\rho$ meson compared to the $\pi$.

\begin{figure}
\unitlength1.cm
\begin{picture}(15,9)(1,2.5)
\includegraphics{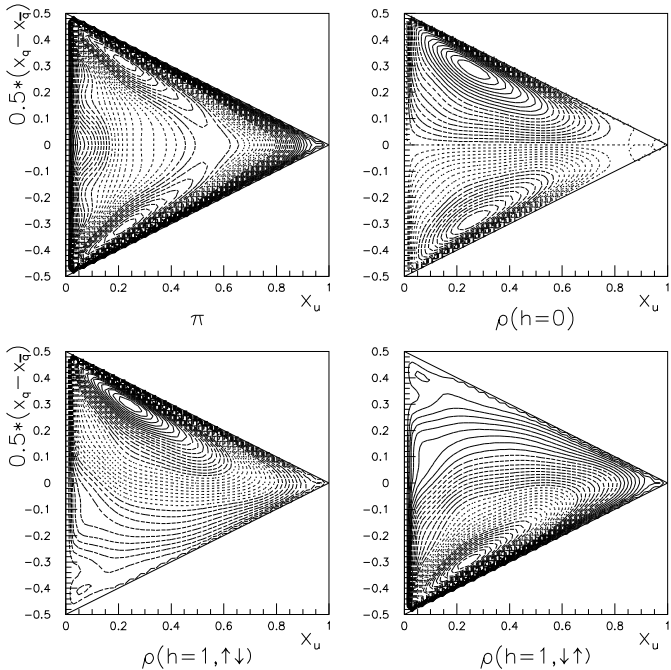}
\end{picture}
\caption{LF wave functions for the $\pi$, and the $h=0,1$ states of the
$\rho$ meson in the Fock component with $q\bar{q}$ and one link field $U$ and
with helicity for the quarks flipped compared to the valence component. 
The x-axis is the momentum fraction $x_U$ carried by the link-field and the
y-axis is the difference between the momentum fraction carried by the quark
and the anti-quark $x_q-x_{\bar{q}}$. For the $\rho$ meson with $h=1$
there are two
possible spin configurations for the quarks, which have different wave
functions.}
\label{fig:wavef3v}
\end{figure}

\begin{figure}
\unitlength1.cm
\begin{picture}(15,5.5)(.7,3.5)
\includegraphics{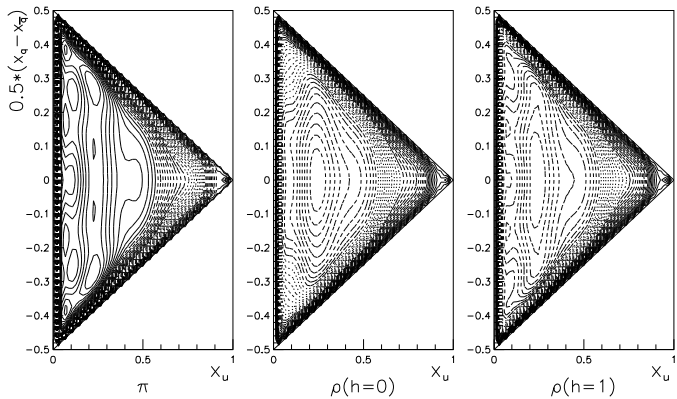}
\end{picture}
\caption{Same as Fig. \ref{fig:wavef3v}, but for the 3-particle component with
the same quark spin wave function as the valence component.}
\label{fig:wavef3s}
\end{figure}

In Figs. \ref{fig:wavef3v}  and \ref{fig:wavef3s} we show the longitudinal
wave functions of the $\pi$ and the $\rho$ in the 3-particle Fock components.
Fig. \ref{fig:wavef3v} shows the component of the 3-particle Fock component where
the quark or the anti-quark helicity has flipped compared to the valence
component (generated by naive hopping!), while Fig. (\ref{fig:wavef3s}) shows
the component where the $q\bar{q}$ helicity wave function is the same as in
the valence state.

\begin{figure}
\unitlength1.cm
\begin{picture}(15,9)(1,2.5)
\includegraphics{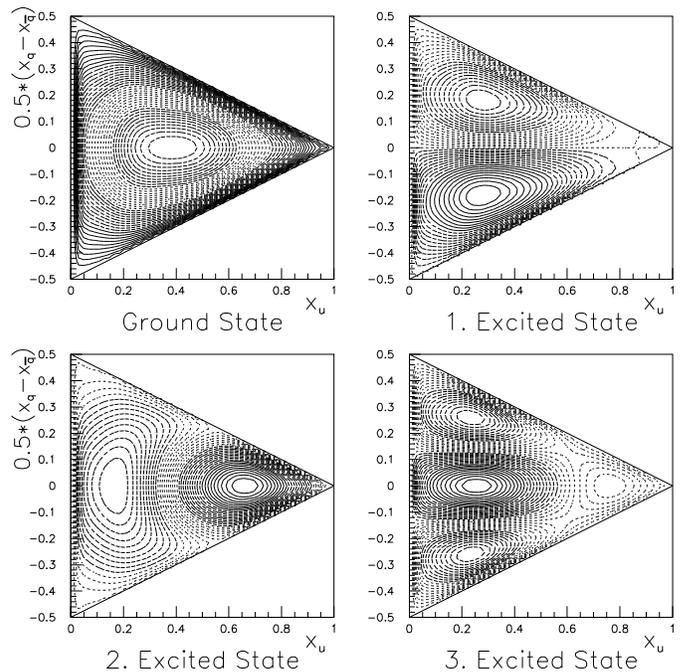}
\end{picture}
\caption{LF wave functions for states that have vanishing $\bar{q}q$
Fock component.}
\label{fig:hybrid}
\end{figure}

In Section \ref{sec:quq} we demonstrated that there are states in the
spectrum which do not mix with the $\bar{q}q$ component of the Fock space.
In Fig. \ref{fig:hybrid} we show the (longitudinal) LF wave functions for
the four lightest solutions to Eq. (\ref{eq:quq}). The lightest solution to
Eq. (\ref{eq:quq}), displayed in the upper left of Fig. \ref{fig:hybrid},
has no nodes, i.e. we interpret this solution as a $\bar{q}q$ state with
orbital angular momentum around the z-axis (see the discussion in Section
\ref{sec:quq}) but with the glue in its ground state. The mass of this
state is about 1.63 GeV, i.e. much higher than experimentally observed mesons
with $J=2$, which we attribute to the one link approximation.
The second excited solution to the $\bar{q}Uq$ equation (lower left in Fig.
\ref{fig:hybrid}) has a node in the wave
function for the link field relative to the $\bar{q}q$ pair, i.e. this
state should be interpreted as a hybrid excitation. Its mass is about 
$2.27 GeV$, but again we expect significant corrections from higher Fock
components which we have not included. 
Finally, the first and third excited state (upper and lower right in
Fig. \ref{fig:hybrid} respectively) correspond to states where there
are nodes in the $\bar{q}q$ relative wave function, but not relative to
the link field, i.e. we interpret these solutions again as non-hybrid
excited states. Their masses are $2.01 GeV$ and $2.32 GeV$ respectively.

\begin{figure}
\unitlength1.cm
\begin{picture}(15,9)(1,2.5)
\includegraphics{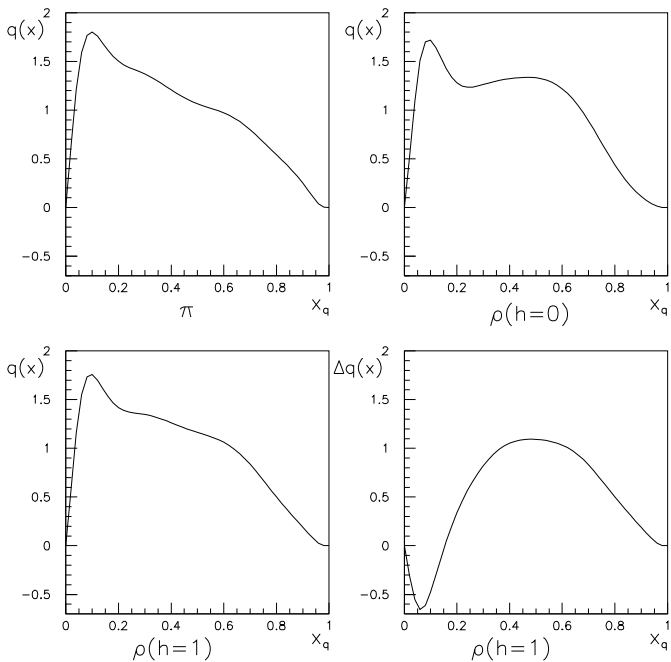}
\end{picture}
\caption{Unpolarized quark distribution functions for the $\pi$, the
$\rho$ with helicity $h=0$ and the $\rho$ with $h=1$. For the $\rho$ with
$h=1$ the polarized quark distribution is also shown.}
\label{fig:stru}
\end{figure}

Quark distribution functions (``structure functions'') for the $\pi$ and
$\rho$ mesons are shown in Fig. \ref{fig:stru}. For the $\rho$ with
helicity $h=1$, the polarized quark distribution function is also displayed.
Note that, since the $\rho$ meson has spin 1, even its unpolarized quark
distribution function does not need to be the same for the h=0 and h=1
states, i.e. the fact that our numerical results for the unpolarized
distribution functions of the $\rho$ states with $h=0$ and $h=1$ should
not necessarily be interpreted as a manifestation of rotational invariance
violations! We should emphasize that the quark distribution functions
should {\it not} vanish at the origin, because the interaction coupling
the 2 particle and 3 particle Fock components is singular enough as the
fermion momentum goes to zero to overcome the suppression due to the
rise of the kinetic energy in the 3 particle component when the
momentum of one of the particles vanishes. However, in the numerical
results this singularity is suppressed due to our cutoff on the
momentum fractions in a hopping (i.e. link field emission) process.
Varying this cutoff $\varepsilon$ hardly changes the structure functions except
in the extreme vicinity of $x=0$, where the ``gap'' fills up as 
$\varepsilon$ is lowered. The structure functions displayed are for
a value of $\varepsilon = 0.1$. The model yields for the momentum fraction 
carried by the quarks plus anti-quarks in the $\pi$ about 
75.4 \%, while
the numbers for the $\rho$ are slightly higher with
$\langle x_q \rangle_{\rho (h=0)} \approx 77.4 \% $ and 
$\langle x_q \rangle_{\rho (h=1)} \approx 76.2 \% $, which again reflects the
stronger binding in the $\pi$ (resulting in a ``smaller'' object, where less
volume is filled with gluon fields).  
In deep inelastic scattering experiments, the structure function of the 
$\pi$ is usually determined at rather large values of $Q^2$. if we were to 
evolve our distribution functions to larger values of $Q^2$ we would
obtain momentum fractions carried by the quarks which are comparable to the
experimentally measured value of about $50 \%$. However, since the
perturbative limit of the $\perp$ lattice is still poorly understood
\cite{bvds}, we will not attempt to evolve our numerical results to
higher values of $Q^2$.

In the polarized quark distribution function for the $\rho$, one can notice
a sign change for small $x$. The reason for this sign change is the following.
At large $x$ the valence quarks dominate, which have the same helicity
as the parent $\rho$). At small $x$, the distribution function is dominated
by quarks (and anti-quarks) which have just emitted a link field and
link filed emission via the naive term leads to helicity flip for the 
quark (or anti-quark). Since the coupling of the naive hopping term is taken
to be larger than the coupling of the r-term, helicity flip dominates at
small $x$ over non-flip and thus the polarized parton distribution changes
sign. In order to interpret this result, we emphasize again that the
parton distributions calculated should be interpreted at a very low momentum
scale. Perturbative evolution would clearly cover up this sign reversal.
However, it is interesting to note that, based on the above results, 
the spin fraction carried by the quarks spin in the $\rho$ meson turns out
to be about 52.9\% . Because of the large $N_C$ approximation employed in
this model, this number should be interpreted as the non-singlet spin
fraction carried by the quarks. Unfortunately, no experimental data on the
structure function of the $\rho$ meson is available, but a number which
has a similar physical interpretation would be the ratio between $g_A$
in the nucleon and its naive quark model value, which turns out to be
$\frac{3}{5}g_A \approx 0.77$. We believe that the relatively small fraction
of spin carried by the quark spins is an artifact of the constant (i.e. 
non-running) dynamical vertex mass which we have used for simplicity
and we expect that a more refined treatment leads to less negative
polarization of the quarks at small $x$.

\section{summary}
One of the main message of this paper is that even though it is possible
to write down a mass term that is invariant under chiral
transformations (on the LF), this does not help to solve the
species doubling problem in a way that is chirally invariant in the
usual sense. \footnote{Once again, since LF Hamiltonians are nonlocal
in the longitudinal direction, the Nielsen-Ninomiya theorem 
\cite{nogo} does not apply directly.}

We used the kinetic mass term, which is chirally invariant in the LF approach,
to remove species doubling. However
the relevant 2 boson -- 2 fermion coupling constants are not
independent, but rather must be chosen so that they cancel
divergences arising from iterating the three point interactions.
Once one uses this result, the LF Hamiltonian based on standard
Wilson fermions and the LF Hamiltonian based on a Wilsonian
prescription (applied to the kinetic mass only) become identical. 

As a first application for the transverse lattice formulation of
Wilson fermions, we studied the meson spectrum in a simple model
where we truncated the Fock space to $\bar{q}q$ states ($q$ and $\bar{q}$
on the same site) and $\bar{q}Uq$ states ($q$ and $\bar{q}$ on adjacent sites)
In this approximation, we are able to separate the transverse 
motion, and, for $P_\perp^{total} =0$, we derive an effective collinear
equation for the model which can be easily solved numerically.
After fitting the $\pi$ and $\rho$ masses as well as the longitudinal
string tension, we obtain a qualitatively reasonable meson spectrum,
although excitation energies tend to be systematically too high.
For the lowest hybrid meson state we obtain a mass of $2.27 GeV$, which
we believe is too high due to our Fock space truncation.
The distribution amplitudes for the $\pi$ and $\rho$ mesons are both
peaked at $x=0.5$ and vanishing at the end-points, with the $\rho$ meson
distribution function being slightly more peaked.
The momentum fraction carried by the quarks (and anti-quarks) in the $\pi$
meson is about $75\% $ with its value for the $\rho$ meson being slightly
higher. For the $\rho$ meson with helicity $h=1$ we find that the spin
fraction due to the spins of the quarks is about 53\% .

Considering the simplicity of the model calculations, the numerical
results on hadron observables are very promising and encouraging. 
However, because of the simplistic nature of our first study
calculations, there are many improvements possible.
First of all, because of the large lattice spacing employed in current
$\perp$ lattice calculations, one might also 
consider introducing ``improvement''
terms \cite{bvds} for quark fields. 
Secondly, especially for studying excited mesons,
it would be very useful to include states with more than one link field
in the Fock expansion, thus allowing the mesons to ``expand'' in the
transverse direction. In connection with including higher Fock components,
one also needs to address the question of dynamical vertex mass generation
\cite{mb:chiral} and the running of the effective vertex mass in more detail 
in order to properly incorporate the physics of dynamical chiral symmetry 
breaking into the effective LF Hamiltonian for QCD on a $\perp$ lattice.

\acknowledgments
We thank Bob Klindworth for critically reading and commenting on
a preliminary version of this paper.
This work was supported by the D.O.E. (grant no. DE-FG03-96ER40965)
and in part by TJNAF.
\appendix
\section{Chiral Symmetry on the LF}
Chiral symmetry on the LF and chiral symmetry in the covariant approach
are not exactly the same (here we closely follow Ref. \cite{osu:all}).
In LF field theory, it is convenient to decompose
$\psi = \psi_+ +\psi_-$, where $\psi_\pm = \frac{1}{2} \gamma^0 
\gamma^\pm \psi$ \footnote{In the main text, we used $\phi \equiv
\psi_+$ and $\chi \equiv \psi_-$ to avoid the use of subscripts.}
and to eliminate $\psi_-$ using a constraint
equation
\begin{equation}
\psi_-=\frac{1}{i\partial^+} \left[ {\vec \alpha}_\perp
\left( i{\vec \partial}_\perp + g {\vec A}_\perp \right)
+ \gamma^0m\right] \psi_+ ,
\label{eq:elim}
\end{equation}
where $ i{\vec D}_\perp \equiv
i{\vec \partial}_\perp + g {\vec A}_\perp$.
Once $\psi_-$ has been eliminated, chiral transformations are applied
to $\psi_+$ only! The constraint equation (\ref{eq:elim}) is
in general inconsistent with the usual chiral transformation, since
\bea
\gamma_5\frac{1}{i\partial^+} \left[ {\vec \alpha}_\perp i{\vec D}_\perp 
+ \gamma^0m\right] \psi_+ &\neq&
\frac{1}{i\partial^+} \left[ {\vec \alpha}_\perp i{\vec D}_\perp
+ \gamma^0m\right] \gamma_5 \psi_+  \nonumber\\
\eea
This explains how it is possible that the standard Wilson r-term, which breaks the usual chiral symmetry, does not break LF chiral
symmetry.

It is interesting to note that only for $m=0$ do chiral 
transformations on the LF and those
defined covariantly agree with one another. 
In fact, it turns out that the generator for chiral transformations
on the LF is nothing but the helicity 
of the fermions. This makes sense since for $m=0$ helicity and chirality
become the same (as do chiral transformations defined covariantly
and on the LF) and both become conserved. Given a LF Hamiltonian,
this fact makes it very easy to verify whether it is chirally invariant:
if quark helicity is conserved by each interaction then the Hamiltonian
is chirally invariant.

Manifest (LF-) chiral symmetry implies that the helicity of each quark is 
conserved. This fact gives rise to peculiar degeneracies: for example,
mesons with $S_z=0$ and $C=\pm 1$ (the $\pi$ and the $S_z=0$ polarization states of the $\rho$) are exactly degenerate --- a result which is
phenomenologically not very desirable. Simple estimates show it is
conceivable that a finite $\pi-\rho$ splitting remains in the chiral 
limit due to the singular end-point behavior of wave functions and 
matrix elements of the relevant interactions. However, this point 
requires further clarification.

\end{document}